\documentclass[sigconf]{acmart}
\setcopyright{none}
\renewcommand\footnotetextcopyrightpermission[1]{}
\acmConference[AgenticSE @ KDD 2026]{AgenticSE @ KDD 2026}{2026}{Jeju, Republic of Korea}

\usepackage{amsmath}

\usepackage{amssymb}
\usepackage{multirow}
\usepackage{booktabs}
\usepackage{colortbl}
\usepackage{relsize}
\usepackage{graphicx}
\usepackage{tikz}
\newcommand{\blackcircle}[1]{%
  \begin{tikzpicture}[baseline=(char.base)]
    \node[shape=circle, draw, inner sep=1.2pt, fill=black, text=white, font=\bfseries\small] (char) {#1};
  \end{tikzpicture}%
}
\usepackage{cleveref}
\crefname{table}{Tab.}{Tabs.}
\Crefname{table}{Table}{Tables}
\crefname{figure}{Fig.}{Figs.}
\Crefname{figure}{Figure}{Figures}
\crefname{equation}{Eq.}{Eqs.}
\Crefname{equation}{Equation}{Equations}
\crefname{section}{Sec.}{Secs.}
\Crefname{section}{Section}{Sections}
\crefname{subsection}{Subsec.}{Subsecs.}
\Crefname{subsection}{Subsection}{Subsections}

\begin{document}

\title{SVRepair: Structured Visual Reasoning for Automated Program Repair}

\author{Jincheng Wang}
\authornote{Jincheng Wang and Liwei Luo contributed equally to this work.}
\email{xiangjiang.wjc@antgroup.com}
\affiliation{%
  \institution{Ant Group}
  \city{Beijing}
  \country{China}}

\author{Liwei Luo}
\authornotemark[1]
\email{luoliwei.llw@antgroup.com}
\affiliation{%
  \institution{Ant Group}
  \city{Beijing}
  \country{China}}

\author{Xiaoxuan Tang}
\email{leahxx1226@gmail.com}
\affiliation{%
  \city{Beijing}
  \country{China}}

\author{Jingxuan Xu}
\email{xujingxuan.xjx@antgroup.com}
\affiliation{%
  \institution{Ant Group}
  \city{Beijing}
  \country{China}}

\author{Sheng Zhou}
\authornote{Corresponding authors.}
\email{zhousheng\_zju@zju.edu.cn}
\affiliation{%
  \institution{Zhejiang Key Laboratory of Accessible Perception and Intelligent Systems, Zhejiang University}
  \city{Hangzhou}
  \country{China}}

\author{Dajun Chen}
\email{chendajun.cdj@antgroup.com}
\affiliation{%
  \institution{Ant Group}
  \city{Beijing}
  \country{China}}

\author{Wei Jiang}
\email{jonny.jw@antgroup.com}
\affiliation{%
  \institution{Ant Group}
  \city{Beijing}
  \country{China}}

\author{Yong Li}
\authornotemark[2]
\email{liyong.liy@antgroup.com}
\affiliation{%
  \institution{Ant Group}
  \city{Beijing}
  \country{China}}

\renewcommand{\shortauthors}{Wang, Luo, et al.}

\begin{abstract}

Large language models (LLMs) have recently been applied to Automated Program Repair (APR), yet most existing approaches remain unimodal and fail to use diagnostic signals contained in visual artifacts such as screenshots and control-flow graphs.
In practice, many bug reports convey critical information visually (e.g., layout breakage or missing widgets), but directly using such dense visual inputs often causes context loss and noise, making it difficult for MLLMs to ground visual observations into precise fault localization and executable patches.
To bridge this semantic gap, we propose \textbf{SVRepair}, a multimodal APR framework with Structured Visual Representation (SVR).
SVRepair first fine-tunes a vision-language model, SVR, to uniformly transform heterogeneous visual artifacts into a \emph{semantic scene graph} that captures GUI elements and their structural relations (e.g., hierarchy), providing normalized, code-relevant context for downstream repair.
Building on the graph, SVRepair drives a coding agent to localize faults and synthesize patches, and further introduces an iterative visual-artifact segmentation strategy that progressively narrows the input to bug-centered regions to suppress irrelevant context and reduce hallucinations.
Across primary repository-level APR benchmarks, SVRepair resolves \textbf{186/517} SWE-Bench M instances (\textbf{35.98\%} over all instances; \textbf{36.47\%} over submitted runs) and \textbf{4/19} visual OmniGIRL instances (\textbf{21.05\%}).
On supplementary structured multimodal code reasoning benchmarks, SVRepair reaches \textbf{38.02\%} on MMCode and \textbf{95.73\%} on CodeVision.
Code is available at \url{https://github.com/codefuse-ai/CodeFuse-SVR}.

\end{abstract}



\maketitle

\section{Introduction}\label{sec:intro}
\begin{figure}
    \centering
    \includegraphics[width=\columnwidth]{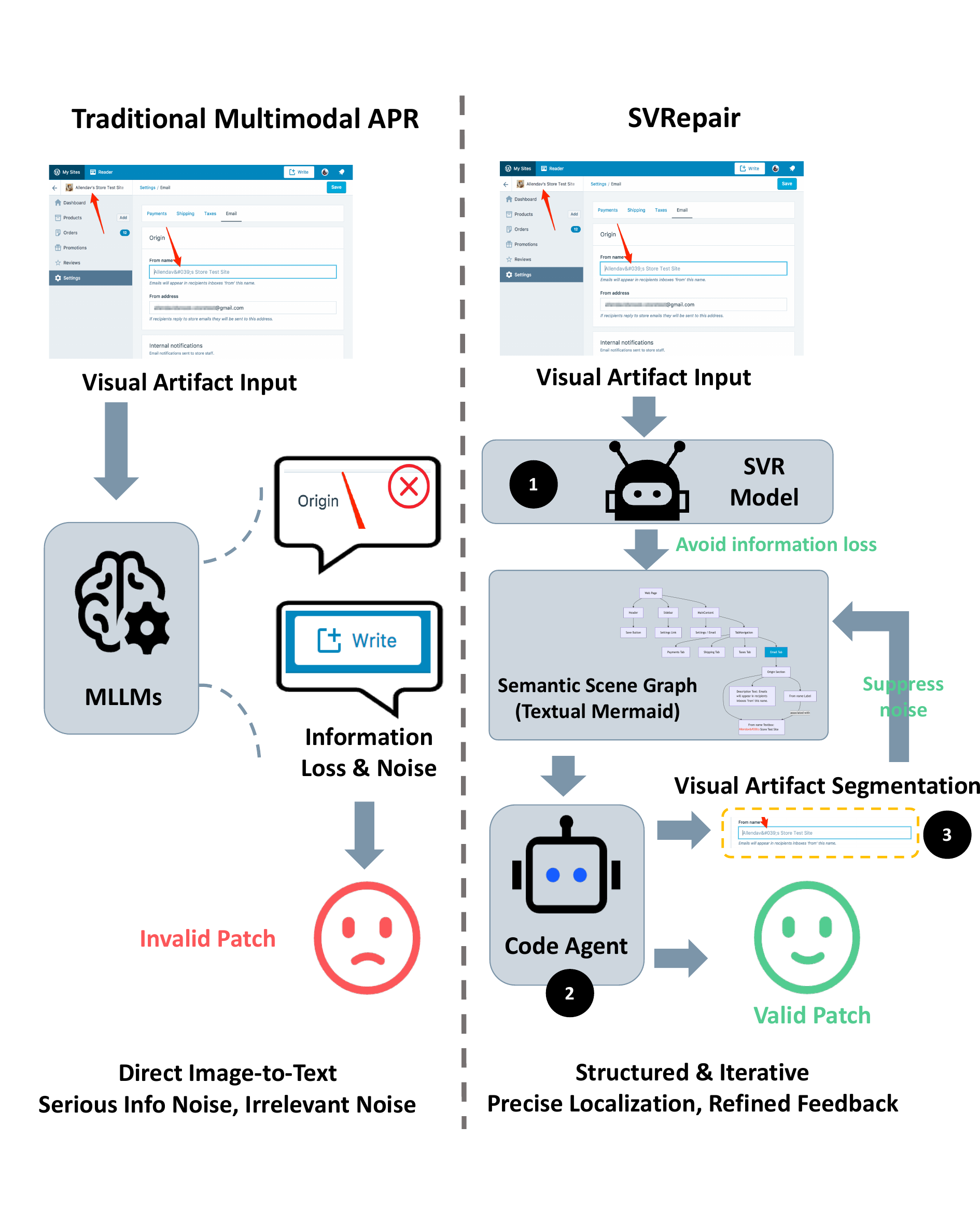}
    \caption{Comparison between traditional Multimodal APR and the proposed SVRepair framework.}
    \Description{A side-by-side workflow comparison showing traditional multimodal APR losing visual context and SVRepair converting visual artifacts into semantic scene graphs before patch generation.}
    \label{fig:motivating-example}
\end{figure}
Automated Program Repair (APR) aims to streamline software maintenance by automatically fixing bugs.
By doing so, software developers minimize manual labor and enhance code reliability~\cite{renzullo2025automated, huang2024evolving}.
Recent LLM-based APR methods use natural-language reasoning and code understanding to localize faults and generate patches~\cite{yang2024swe,ruan2024specrover}.
These methods typically rely on unimodal inputs, such as textual issue reports and specifications.

However, in modern software development, defects are often identified and reported through \textit{visual artifacts} (e.g., screenshots of erroneous web pages and control-flow graphs), which are not captured by unimodal formulations.
A more fundamental obstacle is that many bug reports convey crucial diagnostic signals visually rather than purely through text (e.g., layout breakage, missing widgets, or incorrect rendering states). While modern multimodal LLMs are increasingly capable of perceiving and describing such visual artifacts, they often struggle to ground these observations into the codebase—i.e., to identify the responsible program locations and synthesize correct, executable edits. This mismatch between visual understanding and code-level repair creates a semantic gap that remains a core hurdle for multimodal APR.

Specifically, we identify two primary challenges that hinder the effectiveness of existing APR tools.
The \textbf{first} challenge is \textit{context loss in visual artifacts}. Visual artifacts associated with coding issues often encode fine-grained information about graphical elements and their hierarchical organization. Such context provides strong cues for both fault localization (where the bug manifests) and defect characterization (what kind of bug it is).
For instance, the artifact in Figure~\ref{fig:motivating-example} shows the problematic ``FromName" field and its garbling issue.
Moreover, the element hierarchy relationship reveals that the field is nested within the "Origin" container, directly linking the component structure to the erroneous source files (e.g., notifications-origin.js).
Unfortunately, without extracting these contexts from the visual artifact, it is difficult for APR tools to perform bug localization and patch generation.

The \textbf{second} challenge arises in the density of visual information.
Specifically, since modern software interfaces are dense (e.g., a single screenshot may contain dozens of nested components and state indicators), the visual artifact can include both useful diagnostic evidence and a large amount of bug-irrelevant information (e.g., the upper-right write button in the example artifact).
When the noisy contexts are fed to the LLM, it may hallucinate about the bug location and the patching plan.
Therefore, it is necessary to scope the visual context so that the LLM can localize the code regions to patch.

We propose {SVRepair}, a multimodal APR framework with structured visual representation.
\blackcircle{1} As shown in Figure~\ref{fig:motivating-example}, SVRepair first fine-tunes a vision-language model, which we term Structured Visual Representation. SVR transforms visual artifacts into semantic scene graphs.
This graph records GUI elements and their relations (e.g., hierarchy) in a normalized textual structure, reducing irrelevant context for subsequent LLM processing.
\blackcircle{2} Taking the graph as input, SVRepair drives a coding agent, centered on coding LLMs, to perform bug localization and patch generation.
\blackcircle{3} To mitigate noise from redundant or irrelevant visual context, we further introduce a visual-artifact segmentation strategy that leverages the patch generated in the previous round to refine and segment the visual inputs for subsequent iterations.
The generated sub-artifact will be narrowed to a smaller bug-centered region, and in the next round, it will be fed to SVR to extract more related bug contexts.

We conduct experiments on repository-level APR benchmarks and supplementary code-reasoning benchmarks to evaluate both repair effectiveness and the contribution of structured visual representation. 
Overall, our primary contributions are summarized as follows:
\begin{itemize}
    \item We introduce SVRepair, a multimodal APR framework with structured visual representation.
    The fine-tuned vision-language model maps visual artifacts about coding issues to semantic scene graphs, bridging visual semantics and source code.
    \item To handle complex visual artifacts and redundant visual contexts, we propose a visual segmentation technique to iteratively narrow the artifact into bug-centered regions and improve context extraction for patch generation.
    \item On primary repository-level APR tasks, SVRepair resolves 186/517 SWE-Bench M instances (35.98\% over all instances; 36.47\% over submitted runs) and 4/19 OmniGIRL visual-subset instances (21.05\%). On supplementary code reasoning benchmarks, it achieves 38.02\% on MMCode and 95.73\% on CodeVision.
\end{itemize}

\section{Related Works}\label{sec:related-work}
\subsection{Multimodal Code Generation}

MLLM-based code generation studies how to synthesize executable code or structured markup from visual inputs. In the web/UI setting, prior work develops image-to-HTML generation datasets and evaluations (e.g., Pix2Code~\cite{beltramelli2018pix2code}, WebSight~\cite{laurenccon2024unlocking}, Design2Code~\cite{si2024design2code}) and scales them up with larger webpage-to-code corpora (e.g., Web2Code~\cite{yun2024web2code}, WebCode2M~\cite{gui2025webcode2m}), while some methods incorporate layout-aware modeling to improve structural correctness. In the chart and scientific-plot domain, benchmarks and datasets~\cite{wu2025plot2code, zhao2025chartcoder} evaluate both understanding and chart-to-code reproduction. Related efforts~\cite{wang2025mathcoder} extend to diagram-to-LaTeX conversion and structured vector graphics generation, covering tasks such as converting scientific figures into LaTeX code and producing SVG programs~\cite{yang2025omnisvg, rodriguez2025starvector} for icons and illustrations. General-purpose benchmarks~\cite{li2024mmcode, zhang2025codev} further evaluate multimodal coding with visual inputs in algorithmic problem solving. However, many methods remain task-specific and still struggle with both code executability and faithful visual--code alignment.

\subsection{LLMs/MLLMs for APR}
Researchers have extensively explored LLMs for APR tasks.
Early work addresses software defects through fine-tuning~\cite{jiang2023impact,xia2023automated,wu2023effective} and prompting~\cite{xia2022less,fan2023automated,zhao2024enhancing}.
They usually focus on function-level APR, i.e., analyzing and patching a single function.
Recently, researchers have started exploring agent systems to solve complex repo-level issue scenarios~\cite{zhang2024autocoderover,yang2024swe,xia2025demystifying,ma2025alibaba,antoniades2024swe}.
For instance, SWE-Agent~\cite{yang2024swe} focuses on SWE-Bench~\cite{jimenez2023swe}, and solves SWE problems by designing an Agent Computer Interface to interact with the environment.
While they achieve promising results, they mainly focus on unimodal APR tasks and cannot process repo issues with visual evidence.

A notable contribution to multimodal APR is the recently proposed GUIRepair~\cite{huang2025seeing}.
It uses pre-trained vision-language models (e.g., GPT-4o) to generate issue-reproduction scripts from visual artifacts.
These reproduced scripts are then used to localize buggy files and specific lines of code within the target codebase.
However, this reproduction process often loses critical context from the original figures, such as bug types and related UI components.
Consequently, this insufficient context hinders both precise bug localization and the generation of effective patching plans.
Our approach overcomes these challenges by transforming visual artifacts into a structured intermediate representation that encompasses element attributes in the visual artifact and their hierarchical relationships.
This structured description not only preserves the complete visual context but also provides explicit logical constraints and interpretability for subsequent patch generation.

\section{Method}\label{sec:svr}
\begin{figure*}
    \centering
    \includegraphics[width=\textwidth]{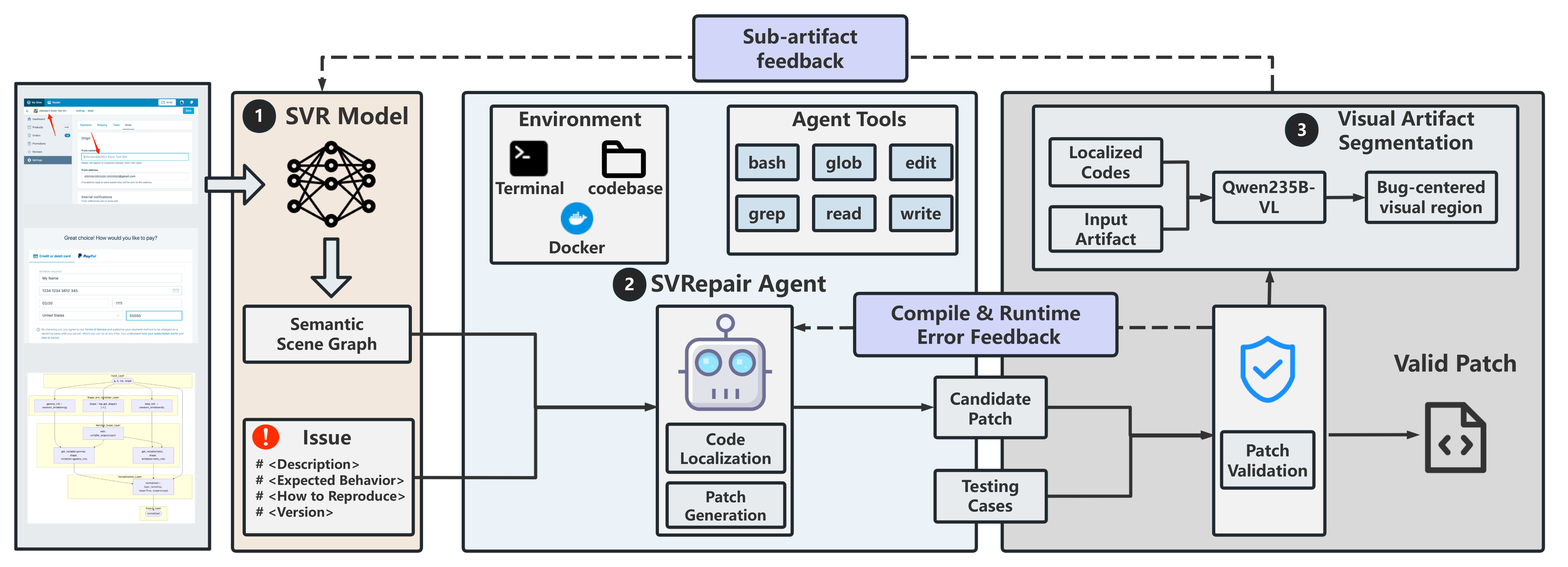}
    \caption{Architecture of SVRepair. The overall workflow comprises three core components: (1) the SVR Model, which transforms input images into semantic scene graphs; (2) the SVRepair Agent, responsible for bug-related code localization and patch generation; and (3) the Visual Artifact Segmentation module, which identifies the most relevant bug-related areas. The SVRepair Agent establishes an iterative refinement loop to produce validated patches that pass test case verification.}
    \Description{SVRepair pipeline showing visual artifact input, SVR graph generation, repository-aware patch generation, validation, and sub-artifact feedback for iterative repair.}
    \label{fig:svrepair-archi}
\end{figure*}

The architecture of SVRepair is illustrated in Figure~\ref{fig:svrepair-archi} and consists of three core modules: (1) the SVR vision-language model for visual artifact reasoning, (2) the coding agent for patch generation, and (3) the patch validation module.
Specifically, SVR processes visual artifacts, e.g., faulty HTML webpage renderings, and produces a structured intermediate representation in textual format (Section~\ref{subsec:svr}).
The coding agent then ingests this IR along with the target code repository to initiate bug localization and patch generation (Section~\ref{subsec:coding-agent}).
Finally, the candidate patches are validated against a suite of predefined test cases.

When processing complex visual artifacts, the initial IR may contain significant noise and coarse-grained bug information.
This lack of precision often leads to overly broad localization results, thereby reducing patching efficiency.
To mitigate this, SVRepair utilizes a feedback loop based on validation results (Section~\ref{subsec:artifact-segmentation}):
It segments complex artifacts into several focused sub-artifacts. The most relevant sub-artifact is then fed back into the SVR model for iterative visual reasoning.
The resulting refined IR provides the granular detail needed to improve subsequent patching rounds.

\subsection{Structured Visual Representation Model}\label{subsec:svr}
\noindent\textbf{IR definition.}
To bridge the semantic gap between heterogeneous visual artifacts (e.g., HTML web pages, flowcharts, and control flow graphs) and executable code, we propose a unified intermediate representation: Semantic Scene Graph (SSG).
This representation serves as a structured guide for SVR training, grounded in the observation that software-centric visual artifacts usually convey information at two distinct levels:
\begin{itemize}
    \item \textit{Element attributes}: The geometric layout, coordinates, and visual appearance of individual components.
    \item \textit{Relational connectivity}: The logical hierarchy and functional relationships between these components.
\end{itemize}
To formalize these observations, we define a Semantic Scene Graph as a directed graph $\mathcal{G} = (\mathcal{V}, \mathcal{E})$.
$\mathcal{V}$ is the set of nodes representing visual elements, e.g., a button element in the HTML webpage or a basic block in a control flow graph.
The set of edges $\mathcal{E}$ represents directional relationships, where an edge $e \in \mathcal{E}$ is a tuple:
$$e = (u, v, r) \quad \text{where} \quad u, v \in \mathcal{V}, \ r \in \mathcal{R}$$
Here, $\mathcal{R}$ denotes a finite set of relation types, including control flow, data flow, and compositional hierarchy.
By capturing both discrete components and their logical interdependencies, the SSG preserves the critical context necessary for comprehensive issue understanding.

To ensure compatibility with downstream coding LLMs, the SSG is serialized into a textual format following Mermaid syntax. Nodes may contain snippets of source code (e.g., HTML tags) or natural language descriptions, while edges utilize Mermaid’s directed edge notation to maintain structural integrity.

\noindent\textbf{Data collection.}
To build the training dataset, we focus on visual artifacts that commonly appear in GitHub issues and multimodal APR benchmarks, including HTML/webpage renderings and program control-flow graphs~\cite{yang2024swe}.
Following the final experimental record, the corpus contains 177k training samples from multiple sources, including WebSight~\cite{laurenccon2024unlocking}, VIRA, McD, TextCaps, PlantUML, and CFG-derived diagram data.
For HTML pages, we transform the raw HTML into the SSG by parsing the document object model (DOM) tree: each HTML element (e.g., \texttt{div} and \texttt{button}) becomes a node, and DOM parent-child relations define composition edges.
For program diagrams, we collect 37 high-rated GitHub repositories based on popularity and issue frequency, extract complex functions, and construct their control-flow graphs with StatiCFG~\cite{staticfg}.
Each CFG node is mapped to an SSG node, and each directed execution edge is mapped to an SSG edge with a control-flow relation type.

\noindent\textbf{Model Training}.
With the collected (visual artifact, SSG) pairs, we train the SVR model to bridge software-centric visual evidence and code-relevant structure.
We use supervised fine-tuning (SFT)~\cite{ouyang2022training} with a standard autoregressive objective:
$$L(\theta) :=-\mathbb{E}_{(x,y)\sim D}\left[\sum_{t=1}^{T} \log P(y_t |x,y_{<t}; \theta)\right],$$
where $D$ is the collected dataset and $(x,y)$ is the (visual artifact, expected SSG) pair.

\subsection{SVRepair Agent}\label{subsec:coding-agent}
Taking the target codebase and the IR as inputs, the SVRepair Agent is responsible for localizing bugs and generating patches.
Towards this end, the agent is equipped with the following capabilities.

\noindent\textbf{Virtual environment setup.}
The agent executes within a secure and isolated Docker~\cite{docker} environment that simulates a real-world development setting.
This environment grants the agent access to the full project codebase, a functional terminal for command execution, and the necessary runtime dependencies (e.g., Node.js, Python) required to build and test the patch.

\noindent\textbf{Tools setup.}
To facilitate interaction with the environment, the agent is equipped with a suite of specialized tools.
\begin{itemize}
    \item \textit{Code navigation tools}: The agent uses grep and glob to perform keyword-based searches across the codebase. It generates search queries based on terms identified in the bug report to narrow down the search space to relevant files.
    \item \textit{Filesystem tools}: Once candidate files are localized, the agent employs read\_file to inspect the source code, and write\_file or edit\_file to apply candidate patches.
    \item \textit{Execution tools}: A bash tool allows the agent to run shell commands, such as installing dependencies, compiling code, or executing test suites. 
\end{itemize}
The agent follows a cyclic Localization $\rightarrow$ Generation $\rightarrow$ Validation workflow.
Specifically, the agent starts by searching the codebase for symbols or error messages mentioned in the IR.
It iteratively reads files to understand the control flow and identify the root cause of the failure.
After identifying the buggy code fragment, the agent leverages a coding LLM to generate a candidate patch.
The prompt provided to the LLM includes the original code context and the issue description, which is shown in Appendix~\ref{app:coding-agent-prompt}.

After patch generation, the agent attempts to verify the fix by running existing test suites (e.g., npm test).
If the environment lacks specific test dependencies (e.g., a missing yarn command or an ERR\_MODULE\_NOT\_FOUND error), the agent can create a standalone validation script (test\_fix.js) using write\_file.
This script mocks the necessary environment variables and imports to verify the logic of the fix in isolation.

\subsection{Visual Artifact Segmentation}\label{subsec:artifact-segmentation}
Once candidate patches are generated, they are validated against a suite of unit test cases.
A patch is considered valid only if it passes all tests.
Otherwise, the validation agent collects feedback (e.g., compilation error logs) to initiate a new round of patch generation.
In our logs, patch success tends to decrease as the complexity of the input visual artifact, measured by the number of elements, increases.

Investigating the root cause, we find that for complex artifacts, SVR generates textual contexts that are comprehensive yet coarse-grained.
Specifically, it records an excessive amount of information regarding bug-irrelevant elements, while the descriptions of bug-relevant elements remain sparse.
For example, in Figure~\ref{fig:motivating-example}, the SVR-generated context includes details about the "Write" button, whereas the actual buggy element (the "FromName" textbox) is only vaguely described.
When these noisy contexts are fed into the SVRepair agent, they hinder precise localization, leading the agent to identify excessively large code regions and produce erroneous patching locations (false positives).
Furthermore, because the bug-relevant information is insufficiently detailed, the agent may misinterpret the bug type and generate incorrect patching plans (false negatives).

To mitigate these challenges, we propose a recursive segmentation strategy for the visual artifacts.
When a candidate patch fails validation, the system extracts a focused sub-artifact centered on the suspected bug region to serve as refined feedback for the subsequent generation cycle.
Specifically, we leverage a pre-trained vision-language model (e.g., Qwen3-VL-235B~\cite{qwen3technicalreport}) to perform precise visual grounding.
By providing the original visual artifact, the issue description, and the localized code snippets as context, we prompt the model to predict the specific coordinates of the bug-relevant area (Appendix~\ref{app:artifact-segmentation-prompt}).
The cropped region filters out irrelevant elements and provides the SVRepair agent with focused context for the next repair round.
To prevent an infinite feedback loop, we implement a maximum iteration threshold.
In our experiments, we set the maximum threshold to three rounds.

\section{Experiments}
In this section, we describe the implementation and evaluation protocol, then report main results, ablations, efficiency measurements, and failure cases.

\subsection{Implementation Details}
For SVR, we use Qwen3-VL-8B~\cite{qwen3technicalreport} as the base VLM for supervised fine-tuning.
Qwen3-VL-8B offers a favorable trade-off between model capability and computational cost, as larger models (e.g., 72B or 235B) are costly to use inside an iterative repair loop.
Following the final experimental record, SVR is trained on 177k samples for 2 epochs with a batch size of 128 and a learning rate of $1.0\times10^{-5}$.
Throughout the paper, we refer to this training stage as supervised fine-tuning (SFT).

\subsection{Evaluation Setup}
\noindent\textbf{Baseline selection.}
We compare SVRepair against two baseline categories.
\begin{itemize}
    \item Multimodal LLMs: Claude~\cite{claude}, Qwen-VL~\cite{bai2025qwen2,qwen3technicalreport}, and GPT-4o~\cite{hurst2024gpt} are selected to establish a foundation for zero-shot cross-modal reasoning.
    \item Autonomous Systems: We evaluate GUIRepair~\cite{huang2025seeing}, Refact Agent~\cite{refactai}, OpenHands~\cite{wang2024openhands}, and Agentless~\cite{xia2024agentless}.
\end{itemize}

\noindent\textbf{Benchmark selection.}
We evaluate on primary repository-level APR benchmarks and supplementary code-reasoning benchmarks.
For primary APR evaluation, we use SWE-Bench M~\cite{yang2024swebenchmultimodal} and the visual subset of OmniGIRL~\cite{guo2025omnigirl}.
SWE-Bench M contains 517 task instances from real JavaScript repositories and evaluates whether autonomous agents can resolve user-facing software engineering issues with visual evidence.
The OmniGIRL visual subset contains 19 multimodal issue-resolution cases with image inputs.
To further test whether SVR translates visual constraints into executable code, we include MMCode~\cite{li2024mmcode} and CodeVision~\cite{wang2025code} as supplementary structured multimodal code reasoning benchmarks.
MMCode contains 3,548 questions and 6,620 images from programming competitions, while CodeVision evaluates code generation from visual flowchart logic.

\subsection{Main Results}
\begin{table*}[t]
\centering
\caption{Pass@1 comparison on primary repository-level APR benchmarks. Counts are shown when available from the experimental records. ``Submitted'' uses successful benchmark submissions as the denominator.}
\label{tab:Swebench-M}
\small
\setlength{\tabcolsep}{5pt}
\begin{tabular}{llccc}
\toprule
\textbf{Method} & \textbf{Base model} & \textbf{SWE-Bench M (all)} & \textbf{SWE-Bench M (submitted)} & \textbf{OmniGIRL visual} \\
\midrule
RAG & GPT-4o & 6.00 & -- & -- \\
SWE-Agent & GPT-4o & 11.99 & -- & -- \\
Agentless Lite & Claude-3.5 Sonnet & 25.34 & -- & -- \\
OpenHands-Versa & Claude-Sonnet 4 & 34.43 & -- & -- \\
\midrule
GUIRepair & GPT-4.1 & 31.14 (161/517) & -- & 0.00 (0/19) \\
SVRepair (Ours) & SVR-8B + GPT-4.1 & \textbf{32.30 (167/517)} & -- & \textbf{5.26 (1/19)} \\
\midrule
GUIRepair & GPT-o3 & 35.98 (186/517) & -- & 0.00 (0/19) \\
\rowcolor[gray]{0.95}
SVRepair (Ours) & SVR-8B + GPT-o3 & 35.98 (186/517) & \textbf{36.47 (186/510)} & \textbf{21.05 (4/19)} \\
\bottomrule
\end{tabular}
\end{table*}

\begin{table}[t]
\centering
\caption{Pass@1 comparison on supplementary structured multimodal code reasoning benchmarks. The Qwen-VL row uses the strongest recorded Qwen-VL baseline for each benchmark.}
\label{tab:mmcode-codevision}
\small
\setlength{\tabcolsep}{8pt}
\begin{tabular}{lcc}
\toprule
\textbf{Method} & \textbf{MMCode} & \textbf{CodeVision} \\
\midrule
GPT-4o & 11.79 & 93.90 \\
Claude 3.5 Sonnet & 27.09 & 82.30 \\
Claude 4.0 & 37.02 & 84.75 \\
Qwen-VL baseline & 34.73 & 88.41 \\
\midrule
\rowcolor[gray]{0.95}
SVRepair & \textbf{38.02} & \textbf{95.73} \\
\bottomrule
\end{tabular}
\end{table}

Table~\ref{tab:Swebench-M} reports the primary APR results.
On SWE-Bench M, SVRepair resolves 186/517 instances, corresponding to 35.98\% over all instances and 36.47\% over submitted runs.
The all-instance point estimate matches GUIRepair under the same GPT-o3 coding model, and the corresponding submitted-run metric is 36.47\%.
Because this margin is small and we do not have paired per-instance significance tests for all baselines, we interpret the SWE-Bench M result as competitive rather than as statistically significant evidence of superiority.
For the GPT-4.1 controlled comparison, SVRepair improves from 31.14\% to 32.30\%, again suggesting that structured visual representation helps under the same coding model.

On OmniGIRL, the visual subset is small but directly tests benchmark transfer to a new issue-resolution environment.
SVRepair with GPT-o3 resolves 4/19 cases (21.05\%; Wilson 95\% CI: 8.51--43.33\%).
GUIRepair resolves 0/19 cases (0.00\%; Wilson 95\% CI upper bound: 16.82\%).
This gap should be read together with the workflow difference: GUIRepair depends on benchmark-specific validation and candidate-selection workflows, while SVRepair lets the coding agent construct and execute tests within the target repository.

Table~\ref{tab:mmcode-codevision} reports supplementary code-reasoning results.
These benchmarks are not repository-level APR tasks; instead, they evaluate whether the structured representation preserves enough visual logic for executable code generation.
SVRepair achieves 38.02\% on MMCode and 95.73\% on CodeVision, providing auxiliary evidence that SSG-style intermediate representations can improve code-oriented visual reasoning.

\subsection{Ablation Study}\label{subsec:ablation-study}
\begin{table*}[t]
\centering
\caption{Ablation study of SVRepair. V4 is evaluated only on SWE-Bench M because sub-artifact feedback targets noisy repository-level screenshots rather than curated code-generation images.}
\label{tab:ablation_with_checks}
\small
\setlength{\tabcolsep}{7pt}
\begin{tabular}{lcccccc}
\toprule
\multirow{2}{*}{\textbf{ID}} & \multicolumn{3}{c}{\textbf{Ablation Design}} & \multicolumn{3}{c}{\textbf{Pass@1 (\%)}} \\
\cmidrule(lr){2-4} \cmidrule(lr){5-7}
& \textbf{Vision} & \textbf{SVR (IR)} & \textbf{Feedback} & \textbf{SWE-Bench M} & \textbf{MMCode} & \textbf{CodeVision} \\
\midrule
(1) & -- & -- & -- & -- & 16.92 & 60.36 \\
(2) & Y & -- & -- & 31.14 & 16.33 & 89.02 \\
(3) & Y & Y & -- & 35.01 & \textbf{38.02} & \textbf{95.73} \\
\midrule
\rowcolor[gray]{0.95}
(4) & Y & Y & Y & \textbf{35.98} & -- & -- \\
\bottomrule
\end{tabular}
\end{table*}

Table~\ref{tab:ablation_with_checks} isolates the contribution of visual input, SVR, and the sub-artifact feedback loop.
For CodeVision, adding visual information improves Pass@1 from 60.36\% to 89.02\%, confirming that flowchart images carry essential program logic beyond the textual prompt.
For MMCode, raw image captions alone do not help, but replacing them with SVR-generated SSGs raises Pass@1 from 16.33\% to 38.02\%.
This contrast supports the central design choice: the downstream coding model benefits more from normalized visual structure than from unstructured natural-language captions.

The feedback loop mainly affects dense repository-level screenshots.
On SWE-Bench M, adding feedback increases the resolved rate from 35.01\% to 35.98\% in the SVR-8B + GPT-o3 setting, while the submitted-run metric reaches 36.47\%.
For successfully resolved SWE-Bench M issues, the average number of repair rounds is 1.8 under a maximum threshold of three rounds, indicating that a non-trivial fraction of repairs require more than a single visual-reasoning pass.
In a K2-based feedback audit, visual feedback improves the resolved count from 169/517 to 172/517; among 162 feedback-triggered submissions, the feedback-enabled log records 8/162 resolved cases, while the no-feedback log records 0/162.
The feedback detector records 85.96\% precision (251/292), 74.70\% recall (251/336), and 79.94\% F1 for identifying unresolved cases that should receive another visual feedback round.

To further evaluate SVR itself, we assess Mermaid diagram parsing.
We measure Rendering Accuracy to verify syntactic validity and SSIM to evaluate structural fidelity by comparing rendered predictions against ground-truth diagrams.
The test set contains approximately 1,300 code-control-flow graph pairs from high-starred GitHub repositories.

\begin{table}[t]
\centering
\caption{Mermaid diagram parsing results. SVR keeps rendering accuracy close to a much larger VLM while improving structural fidelity.}
\label{tab:Mermaid_Diagram}
\small
\begin{tabular}{lcc}
\toprule
\textbf{Model} & \textbf{Rendering Acc. (\%)} & \textbf{SSIM} \\
\midrule
Qwen3-VL-235B & 94.97 & 0.7006 \\
Qwen3-VL-8B & 81.14 & 0.6868 \\
SVR-8B & 94.29 & \textbf{0.7892} \\
\bottomrule
\end{tabular}
\end{table}

As shown in Table~\ref{tab:Mermaid_Diagram}, SVR reaches 94.29\% Rendering Accuracy, close to Qwen3-VL-235B (94.97\%) and higher than the Qwen3-VL-8B base model.
SVR also obtains the highest SSIM (0.7892), suggesting that supervised training on SSG targets improves structural fidelity rather than only syntactic validity.

\subsection{Efficiency and Failure Analysis}
\noindent\textbf{Efficiency characterization.}
The experiment summary logs record 1,219,883.71 prompt tokens and 7,792.91 completion tokens for SVRepair.
The resulting total is 1,227,676.62 tokens, with recorded costs of \$0.71 under Kimi-K2 and \$2.50 under GPT-o3.
The same logs report an average of 56.45 calling rounds and 617.13 seconds per instance.
These measurements are not a fully controlled end-to-end cost benchmark against GUIRepair, because the logs do not disaggregate every token field by benchmark and baseline.
They nevertheless characterize the operational trade-off: SVRepair uses greedy one-patch-per-iteration repair with feedback, whereas GUIRepair samples up to 40 candidate patches per issue for selection.

\noindent\textbf{Failure modes.}
We manually inspect SWE-Bench M logs and group remaining failures into five recurring categories.
First, some patches repeatedly fail validation: runs of \texttt{npm test} return exit code 1, but the captured logs do not always expose the concrete failing assertion.
Second, the agent can over-edit the repository; for example, \texttt{alibaba-fusion\_\_next-101} produces a 26,878-line patch that includes unrelated files such as lockfiles and build artifacts.
Third, tool and environment errors can block localization, including malformed tool calls such as \texttt{GlobTool.execute()} without a pattern and file-access attempts outside the workspace.
Fourth, visual parsing errors remain consequential: the logs contain image-caption failures such as \texttt{NoneType} errors in 16 resolved cases and 43 unresolved cases, and invalid-patch counts of 24 and 37 in the corresponding resolved/unresolved groups.
Finally, large or unusual visual artifacts may still defeat cropping; in \texttt{carbon-5485}, the underlying defect is an oversized image overflowing its display frame, which requires precise reasoning about scale and container boundaries.
These failures motivate the limitations discussed in Section~\ref{sec:limitation}.

\subsection{Case Study}
\begin{figure}
    \centering
    \includegraphics[scale=.27]{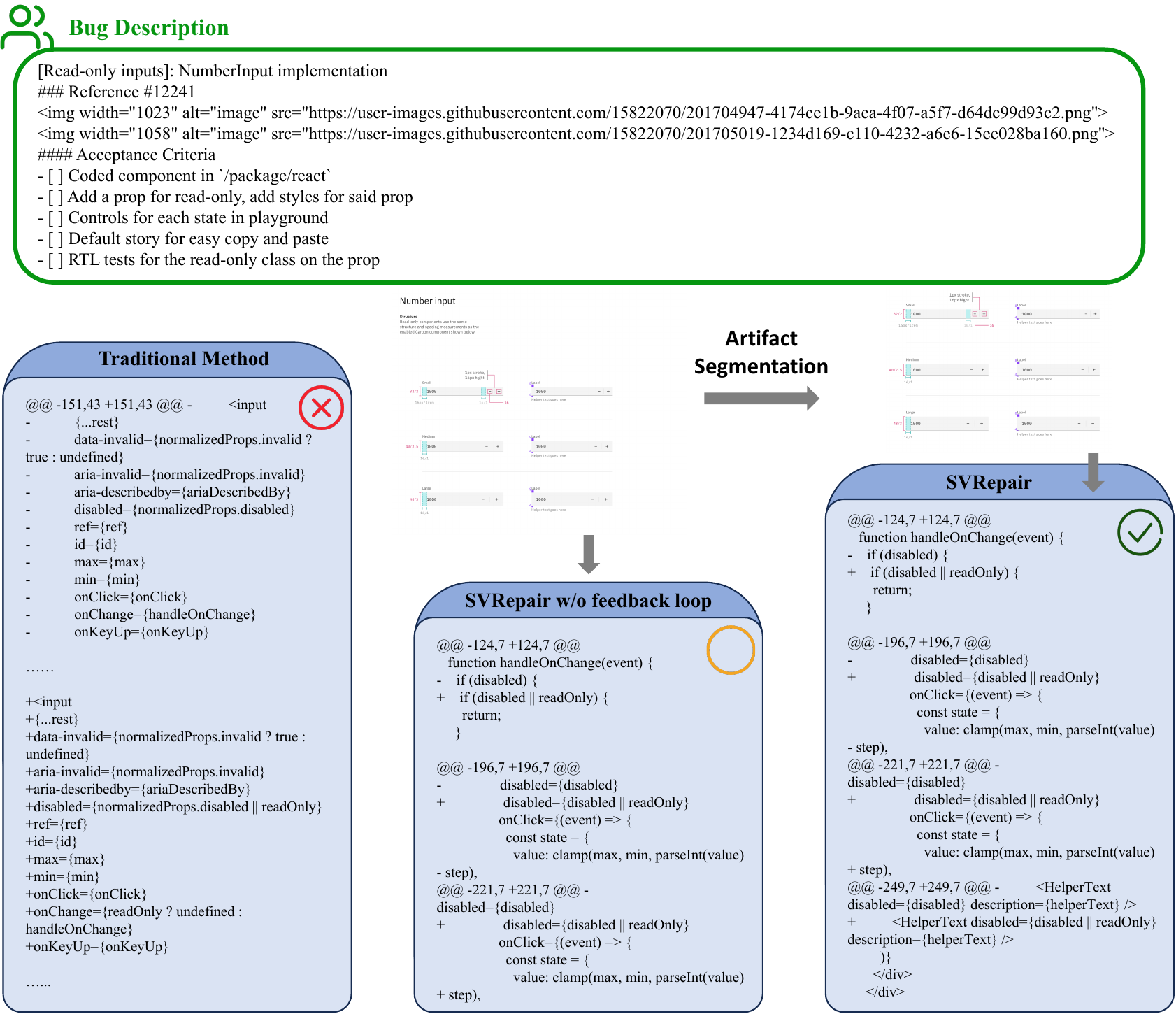}
    \caption{A case study comparing a traditional repair workflow with SVRepair.}
    \Description{A NumberInput repair example in which SVRepair narrows the visual artifact, generates a structured representation, and produces a focused patch compared with a broader traditional workflow.}
    \label{fig:case-study}
\end{figure}

Figure~\ref{fig:case-study} shows a NumberInput component bug that requires read-only functionality under specific acceptance criteria.
The traditional workflow on the left relies on manually defined test paths and broad candidate edits.
It modifies many potentially related properties, including \texttt{data-invalid}, \texttt{aria-invalid}, \texttt{aria-describedby}, \texttt{disabled}, \texttt{ref}, \texttt{id}, \texttt{max}, \texttt{min}, and event handlers, which increases the risk of missing the true root cause or introducing side effects.

SVRepair follows a more constrained path.
The SVR-generated SSG preserves the relationship between the visible NumberInput state, the relevant component hierarchy, and the code locations that govern read-only behavior.
When the first patch fails validation, the feedback loop crops the artifact to the bug-centered region and regenerates a more focused representation.
The coding agent can then edit the read-only control path instead of scattering changes across unrelated properties.
This case illustrates why structured visual context and feedback-based narrowing are complementary: the former provides code-relevant visual semantics, while the latter reduces the noise that accumulates in dense GUI screenshots.

\section{Conclusion}
This paper presents SVRepair, a multimodal automated program repair framework that bridges visual diagnostics and source-code repair through Structured Visual Representation.
SVRepair transforms heterogeneous visual artifacts, such as screenshots and control-flow graphs, into Semantic Scene Graphs that preserve visual elements and their structural relations for downstream coding agents.
Its feedback-guided segmentation further narrows dense screenshots into bug-centered regions when initial patches fail validation.
Across repository-level APR benchmarks and supplementary code-reasoning benchmarks, the results suggest that structured visual representations are a promising way to make visual evidence usable for executable patch generation.
The remaining failures also show that multimodal APR is still bounded by environment reproducibility, test adequacy, visual parsing errors, and the cost of iterative feedback, which motivate future work on more robust validation and broader visual-artifact coverage.

\section{Limitations}\label{sec:limitation}
SVRepair is currently optimized for the visual artifact types covered by our training data and benchmarks, especially HTML and webpage renderings and program control-flow graphs.
Although the Semantic Scene Graph format can in principle represent other software artifacts, such as sequence diagrams, architecture diagrams, or log visualizations, these domains require additional data and validation before the same conclusions can be claimed.

The feedback loop improves dense repository-level screenshots, but it also introduces extra inference and execution cost.
Our logs record an average of 56.45 calling rounds and 617.13 seconds per instance, and the segmentation module relies on a large grounding VLM for crop selection.
We therefore treat sub-artifact feedback as a targeted mechanism for noisy issue screenshots rather than a universally beneficial step; for curated code-generation images in MMCode and CodeVision, applying such cropping may remove useful information.

Finally, SVRepair inherits common limitations of benchmark-based APR.
Passing benchmark tests does not guarantee full semantic correctness, and failing tests may reflect incomplete environment reproduction rather than an incorrect patch.
Docker-based reproduction helps standardize execution, but OS-specific dependencies, missing packages, flaky tests, and incomplete logs can still block validation.
These limitations motivate stronger semantic or human validation in future multimodal APR evaluations.

\begin{acks}
We thank the reviewers for their constructive feedback. TBD.
\end{acks}

\appendix
\section{Prompt Skeletons}
The full prompts used by SVRepair are included in the released repository. This appendix summarizes their structure to clarify the information passed between SVR, the coding agent, and the visual-segmentation module.

\subsection{Coding Agent Prompt}
\label{app:coding-agent-prompt}
The coding agent prompt provides the textual issue report, the SVR-generated Semantic Scene Graph, and the current repository context. It instructs the coding LLM to:
\begin{enumerate}
    \item infer the likely root cause from the issue report and structured visual context;
    \item localize relevant files and code regions using repository navigation tools;
    \item generate a single candidate patch in structured edit blocks;
    \item run available tests or construct a focused validation script when the original test environment is incomplete;
    \item use validation feedback for the next repair iteration if the patch fails.
\end{enumerate}
This structure enforces executable edits and keeps the generated patch aligned with both the visual evidence and repository-level validation.

\subsection{Artifact Segmentation Prompt}
\label{app:artifact-segmentation-prompt}
The segmentation prompt is used only when a candidate patch fails validation, and the visual artifact is likely too dense for precise localization. It provides the original image, the issue description, and localized code snippets from the previous repair attempt. The grounding VLM is asked to identify the bug-relevant visual region and return a bounding box in the form \texttt{[x, y, w, h]}. The cropped sub-artifact is then passed back to SVR to regenerate a more focused Semantic Scene Graph for the next repair round.

\bibliographystyle{ACM-Reference-Format}
\bibliography{reference}

\end{document}